\documentclass[final,5p,times,twocolumn]{elsarticle}

\biboptions{sort&compress}

\usepackage{graphicx}
\usepackage{graphics}
\usepackage{multirow}
\usepackage{multicol}
\usepackage{caption}
\usepackage{subcaption}
\usepackage[hidelinks]{hyperref}

\usepackage{amsfonts}
\usepackage{amsmath}
\usepackage{mathtools}
\usepackage{amssymb}
\DeclareMathAlphabet{\mathcal}{OMS}{cmsy}{m}{n}

\newcommand{\Nf}{{N_f}}
\newcommand{\psibar}{{\overline{\psi}}}

\newcommand{\cZ}{{\cal Z}}
\newcommand{\MSbar}{{\overline{\text{MS}}}}

\newcommand{\bs}[1]{{\boldsymbol{#1}}}

\newcommand{\vxi}{{\bs \xi}}

\usepackage{xcolor}

\begin{document}

\begin{frontmatter}



\title{Equation of State of QCD with $N_f = 3$ flavours up to the electroweak scale}


\author[add1]{Matteo Bresciani\corref{cr}} 
\ead{mbrescia@tcd.ie}
\author[add2,add3]{Mattia Dalla Brida} 
\ead{mattia.dallabrida@unimib.it}
\author[add2,add3]{Leonardo Giusti} 
\ead{leonardo.giusti@unimib.it}
\author[add3]{Michele Pepe} 
\ead{michele.pepe@mib.infn.it}

\cortext[cr]{Speaker}

\address[add1]{School of Mathematics and Hamilton Mathematics Institute, Trinity College Dublin, Dublin 2, Ireland}
\address[add2]{Dipartimento di Fisica, Università di Milano-Bicocca, Piazza della Scienza 3, I-20126 Milano, Italy}
\address[add3]{INFN, Sezione di Milano-Bicocca, Piazza della Scienza 3, I-20126 Milano, Italy}

\begin{abstract}
The Equation of State of Quantum Chromodynamics with $\Nf=3$ flavours is determined non-perturbatively with a precision 
of about $0.5\%-1.0\%$ in the range of temperatures between 3 GeV and 165 GeV. 
The computation is carried out by numerical simulations of the gauge theory discretized on the lattice. 
At each given temperature the entropy density is computed at several lattice spacings in order to extrapolate the results 
to the continuum limit. 
The pressure and energy density are then determined by integrating the entropy density with respect to the temperature. 
The numerical data show a linear behaviour in the strong coupling constant squared, which points to the Stefan-Boltzmann limit 
at infinite temperature. 
They are also compatible with the known perturbative formula supplemented by higher order terms in the coupling constant, 
containing non-perturbative contributions. 
This parametrization describes well our data together with those present in the literature down to 500 MeV.
\end{abstract}

\end{frontmatter}


\section{Introduction}
In the Standard Model, Quantum Chromodynamics (QCD) is the sector which describes the strong interactions among quarks and gluons.
The thermodynamic properties of strongly interacting matter at equilibrium are encoded by the QCD Equation of State (EoS), 
which is of primary interest, for instance, in the study of the evolution of the Early Universe.
The EoS is in fact a crucial component in the determination of the spectrum of primordial Gravitational Waves~\cite{Saikawa:2018rcs},
as well as the abundance of Dark Matter candidates in the Universe~\cite{Saikawa:2020swg,Borsanyi:2016ksw,GrillidiCortona:2015jxo}.
For cosmological purposes, the EoS is needed at all temperatures up to the electroweak scale.
In the literature, the available non-perturbative determinations of the EoS reach temperatures
up to 2 GeV for $\Nf=2+1$ flavours~\cite{Borsanyi:2013bia, HotQCD:2014kol,Bazavov:2017dsy}, 
and up to 1 GeV for $\Nf=2+1+1$ flavours~\cite{Borsanyi:2016ksw}.
At higher temperatures, the standard approach to determine the EoS is by treating QCD perturbatively.
However, the perturbative expansion holds only up to order $g^6$ in the 
strong coupling $g$, where non-perturbative effects start to contribute~\cite{Linde:1980ts,Kajantie:2002wa}.
Furthermore, the poor convergence of the perturbative series up to the 
electroweak scale introduces large systematic uncertainties in the EoS,
which propagate to the cosmological models weakening their predictive 
capabilities~\cite{Saikawa:2018rcs,Saikawa:2020swg}.

In these proceedings we present the non-perturbative determination of 
the EoS in the temperature interval from 3 GeV to 165 GeV~\cite{Bresciani:2025vxw}.
This result has been obtained thanks to a new strategy which overcomes 
some technical limitations of the state of the art method when applied to temperatures above 1 GeV or so.
On the one hand, we abandon hadronic renormalization schemes and we
determine the lines of constant physics by fixing the value of a renormalized
coupling, defined non-perturbatively~\cite{Luscher:1991wu,Luscher:1993gh,Jansen:1995ck,DallaBrida:2021ddx}.
On the other hand, we formulate thermal QCD in the presence of shifted 
boundary conditions~\cite{Giusti:2010bb,Giusti:2011kt,Giusti:2012yj,DallaBrida:2020gux}, 
where the Equation of State can be accessed without 
the need of subtracting explicitly the power-like divergence in the thermodynamic potentials.

For this computation we have chosen the setup of QCD with $N_f=3$ massless quarks. 
The effect of the up, down and strange quark masses is negligible within our final accuracy at the 
temperatures under investigation~\cite{Borsanyi:2016ksw,Bazavov:2017dsy,Laine:2006cp}.
Our primary observable is the entropy density, which we have computed for each 
temperature at several values of the lattice spacing to extrapolate it to the continuum limit.
Pressure and energy density can then be determined by standard thermodynamic relations.
We finally compare our non-perturbative results to the ones in the literature
at lower temperatures, and to the predictions of perturbation theory.

\section{The EoS of QCD from shifted boundary conditions}
In the Euclidean continuum spacetime, thermal QCD can be formulated in a moving reference frame by exploiting 
the soft breaking of the SO$(4)$ invariance of the theory.
In the path integral formalism, this setup is equivalent to consider shifted boundary conditions for the fields
along the temporal direction~\cite{Giusti:2010bb,Giusti:2011kt,Giusti:2012yj,DallaBrida:2020gux}.
The gauge field $A_\mu$, belonging to the algebra of the colour group SU$(3)$, satisfies
\begin{equation}
    A_\mu(x_0+L_0, \bs{x}) = A_\mu(x_0, \bs{x}-L_0\vxi )\,,
\label{eq:shBCs_A}
\end{equation}
where $L_0$ is the size of the temporal direction and $\vxi=(\xi_1,\xi_2,\xi_3)$ is the three-dimensional shift vector.
As matter content, we consider $N_f=3$ flavours of quarks in the fundamental representation of SU$(3)$.
Along the temporal direction, the fermionic fields $\psi,\psibar$ satisfy
\begin{equation}
\begin{aligned}
    \psi(x_0+L_0, \bs{x}) &= -\psi(x_0, \bs{x}-L_0\vxi )\,, \\
    \psibar(x_0+L_0, \bs{x}) &= -\psibar(x_0, \bs{x}-L_0\vxi )\,.
\end{aligned}
\label{eq:shBCs_psi}
\end{equation}
In the three spatial directions, each of length $L$, all the fields satisfy periodic boundary conditions.
The free-energy density $f_\vxi$ is defined as usual,
\begin{equation}
    f_\vxi = -\frac{1}{L_0 L^3}\ln\cZ_\vxi\,,
\end{equation}
where $\cZ_\vxi$ is the Euclidean partition function of the thermal system.
In this framework the temperature is related to $L_0$ by $T^{-1}=L_0\sqrt{1+\vxi^2}$ 
and can be varied also by changing the shift at fixed $L_0$.
The entropy density $s(T)$ can thus be computed from the derivative in the shift of the free-energy density,
\begin{equation}
    \frac{s}{T^3} = -\frac{1}{T^3}\frac{\partial f_\vxi}{\partial T} = 
    \frac{1}{T^4} \frac{1+\vxi^2}{\xi_k}\frac{\partial f_\vxi}{\partial \xi_k}\,.
    \label{eq:entropy_cont}
\end{equation}
The entropy is a physical quantity without the need of any power-like divergence subtraction in the free-energy.
Given the entropy, the pressure $p(T)$ can be obtained as the integral of the entropy in the temperature, 
and the energy density $e(T)$ as $e=Ts-p$.

\section{Lattice theory and renormalization}
We consider QCD regularized on a hypercubic lattice with spacing $a$, $L_0/a$ sites in the temporal direction, 
and $L/a$ sites in each of the three spatial directions.
We discretize the pure gauge action with the standard Wilson plaquette action~\cite{Wilson:1974sk}, while
for the fermionic part we consider $N_f=3$ flavours of massless $O(a)$-improved Wilson 
quarks~\cite{Wilson:1975hf,Sheikholeslami:1985ij,Yamada:2004ja}.
In the temporal direction, the link field $U_\mu(x)\in\rm{SU}(3)$ and the fermionic fields $\psi, \psibar$ satisfy analogous 
shifted boundary conditions as in the continuum, see Eqs.~\eqref{eq:shBCs_A} and~\eqref{eq:shBCs_psi}, with $A_\mu$ replaced by $U_\mu$.
Periodic boundary conditions are imposed on the fields along the three spatial directions.

\begin{table}[t]
	\centering
	\begin{tabular}{|c|c|c|}
		\hline
		$T$  & $T$ (GeV) &  $\bar{g}^2_{\rm SF}(\mu=T\sqrt{2})$\\
		\hline
		$T_0$ &  164.6(5.6) &  1.01636 \\
		$T_1$ &  82.3(2.8)  &  1.11000 \\
		$T_2$ &  51.4(1.7)  &  1.18446 \\
		$T_3$ &  32.8(1.0)  &  1.26569 \\
		$T_4$ &  20.63(63)  &  1.3627  \\   
		$T_5$ &  12.77(37)  &  1.4808  \\
		$T_6$ &  8.03(22)   &  1.6173  \\
		$T_7$ &  4.91(13)   &  1.7943  \\
		$T_8$ &  3.040(78)  &  2.0120  \\
		\hline
	\end{tabular}
	\caption{Second column: physical temperatures considered in this work. 
            Third column: values of the Schr\"odinger functional coupling in $\Nf=3$ QCD at the renormalization scale $\mu=T\sqrt{2}$.}
  \label{tab:T0T8GeV}	
\end{table}
We determine the dependence of the bare coupling $g_0$ on the lattice spacing $a$ by fixing the value of a non-perturbatively defined renormalized coupling.
We consider the Schr\"odinger functional (SF) coupling $\bar{g}^2_{\rm SF}$~\cite{Luscher:1992an}, 
whose running in the continuum limit
for $N_f=3$ QCD has been determined precisely~\cite{Bruno:2017gxd,DallaBrida:2018rfy}.
The renormalization condition imposes that the value of this coupling, at a given lattice spacing and renormalization scale $\mu$, matches
the corresponding value in the continuum~\cite{DallaBrida:2021ddx}:
\begin{equation}
    \bar{g}^2_{\rm SF}(g_0^2, a\mu) = \bar{g}^2_{\rm SF}(\mu)\,, \quad a\mu\ll 1\,.
    \label{eq:SFrencond}
\end{equation}
By choosing $\mu=T\sqrt{2}$, this equation gives rise to a family of equivalent renormalization conditions,
so that for each temperature $T$ the function $g_0(a\mu)$ is determined for values of $a$ at which $T$ can be easily accommodated.
Each temperature can thus be simulated at several lattice spacings, and the continuum limit can be taken with systematic effects under control.
Table~\ref{tab:T0T8GeV} reports the 9 values of temperature we considered in this work, from $3$ GeV up to $165$ GeV. 
The corresponding values of the SF coupling are shown too.

The critical mass $m_{\rm cr}$ at a given $g_0$ and $L_0/a$ is then defined by requiring the PCAC mass to vanish 
in the SF setup. 
We refer to Appendix B of Ref.~\cite{DallaBrida:2021ddx} for the technical details.

\section{Numerical computation}

\begin{figure*}[t]
   \begin{center}
   \begin{minipage}{\columnwidth}
   \includegraphics[width=\textwidth]{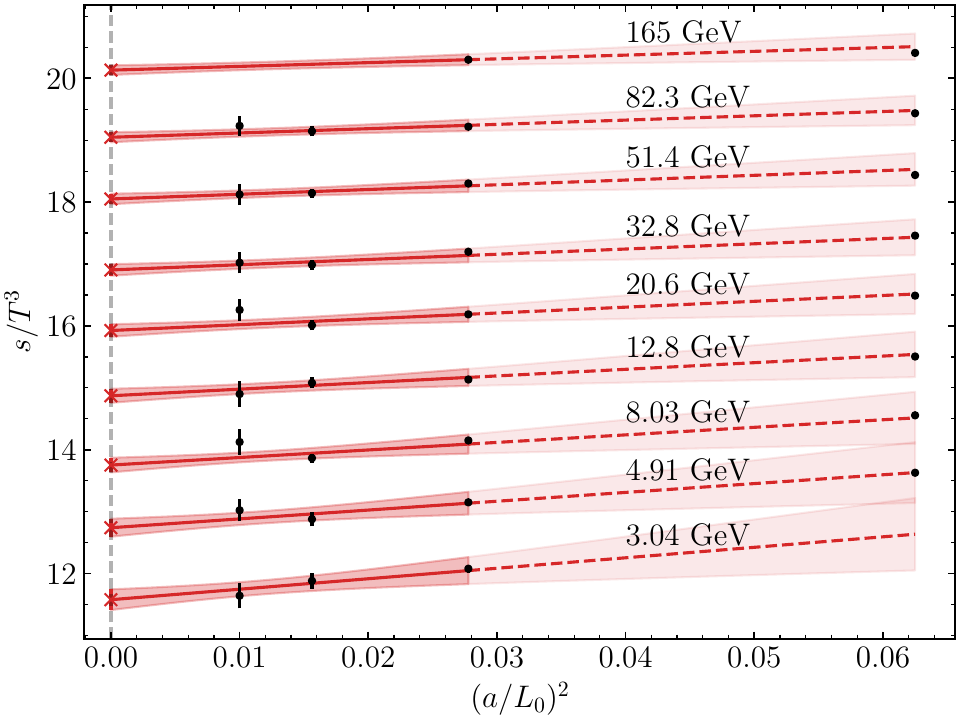}
   \end{minipage}
   \begin{minipage}{\columnwidth}
   \includegraphics[width=\textwidth]{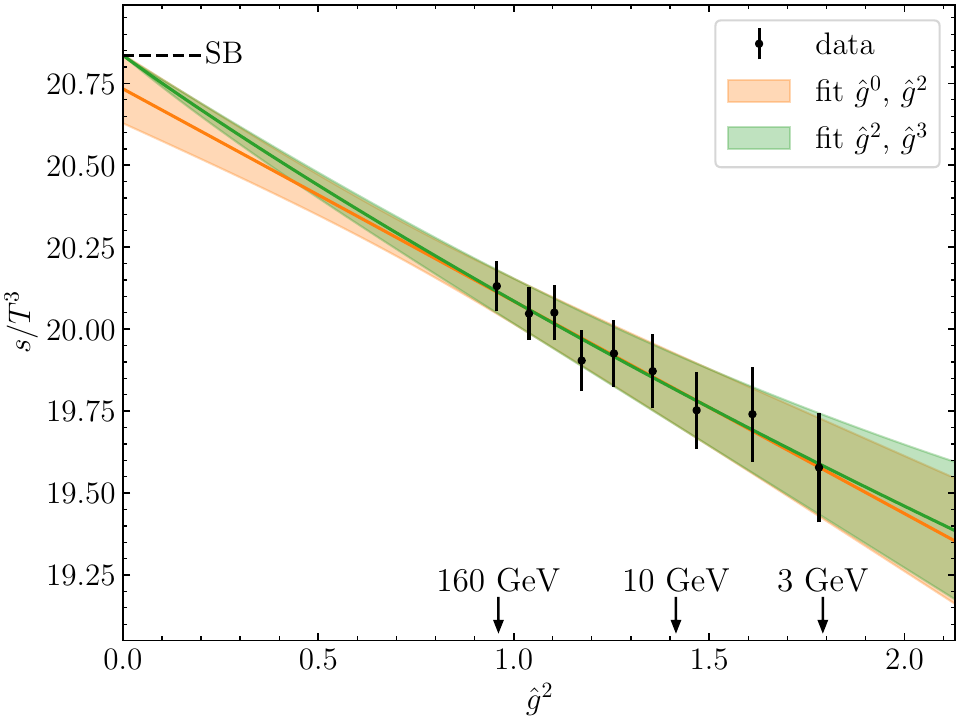}
   \end{minipage}
   \end{center}
   \caption{
   Left: continuum limit extrapolation of $s/T^3$.
   The black points are the one-loop improved non-perturbative results at each given temperature and $a/L_0$, 
   plotted as function of $(a/L_0)^2$.
   Data at each temperature $T_n$, $n=0,1,...,8$ have been shifted downward by $n$ for better readability.
   Red continuum lines and bands represent the continuum limit extrapolation, while red dashed lines and shadowed bands
   are continuations of the fit function for comparison with $L_0/a=4$ data (not included in the fit).
   The continuum limit results are represented by red crosses.
   Right: Continuum results of $s/T^3$ as a function of the coupling $\hat g^2$.
   The orange and green bands are fits to the fit ansatz Eq.~\eqref{eq:fit_ansatz_1}, see main text for details.
   }
   \label{fig:clim_fitpheno}
\end{figure*}
At fixed bare parameters $L_0/a$ and $g_0$, we have chosen the following discretization of the entropy in Eq.~\eqref{eq:entropy_cont}, 
\begin{equation}
    \frac{s}{T^3} = \frac{1+\vxi^2}{\xi_k}\frac{1}{T^4}
    \frac{\Delta f_\vxi}{\Delta\xi_k}\,,
    \label{eq:entropy_lattice}
\end{equation}
where the continuum derivative in $\xi_k$ is replaced by the 2-point symmetric finite difference
\begin{equation}
    \frac{\Delta f_\vxi}{\Delta\xi_k} = 
    \frac{L_0}{4a}\left(f_{\vxi+\frac{2a}{L_0}\hat{k}} - f_{\vxi-\frac{2a}{L_0}\hat{k}}\right)\,,
    \label{eq:f_diff_discrete_shift}
\end{equation}
and $\vxi=(1,0,0)$.
These choices maximize the signal over the statistical noise while keeping discretization effects under control~\cite{inprep}.
The numerical computation on the lattice of the entropy as in Eq.~\eqref{eq:entropy_lattice} 
can be carried out efficiently as described in Refs.~\cite{Bresciani:2025vxw,inprep}.
We have simulated four lattice spacings $L_0/a=4, 6, 8, 10$ in order to take the continuum limit at each 
temperature in Table~\ref{tab:T0T8GeV}.
The spatial volume of our lattices has been sized such that the aspect ratio $LT$ always satisfies the 
constraint $10\lesssim LT\lesssim 25$, and we have also explicitly checked at the highest and lowest temperature that finite volume 
effects are negligible within the final statistical accuracy on the entropy density. 

The non-perturbative results for the entropy density at each given temperature and $L_0/a$ are represented in the left 
panel of Figure~\ref{fig:clim_fitpheno}. 
In order to extrapolate to the continuum limit, we have considered the fit function
\begin{equation}
    s(T, a/L_0)/T^3 = s(T)/T^3 \\+ d_2\,\bar g^3_{\rm SF}(T) \left(\frac{a}{L_0}\right)^2\,,
    \label{eq:fitfunction}
\end{equation}
where the leading discretization effects are of $O(a^2 \bar g^3_{\rm SF})$ because the lattice theory is non-perturbatively $O(a)$-improved,
and cutoff effects up to one-loop order have been subtracted using lattice perturbation theory.
We have performed a global fit of the data at all temperatures including points with $L_0/a>4$.
The fit has a good $\chi^2/\chi^2_{\rm exp}=0.82$, where $\chi^2_{\rm exp}$ is defined in Ref.~\cite{Bruno:2022mfy}, 
and the extrapolated results have a relative error of $0.5\%-1.0\%$.
These results are stable against several different fit procedures~\cite{inprep}.

\section{Equation of State of QCD}

\begin{figure*}[t]
   \begin{center}
   \begin{minipage}{\columnwidth}
   \includegraphics[width=\textwidth]{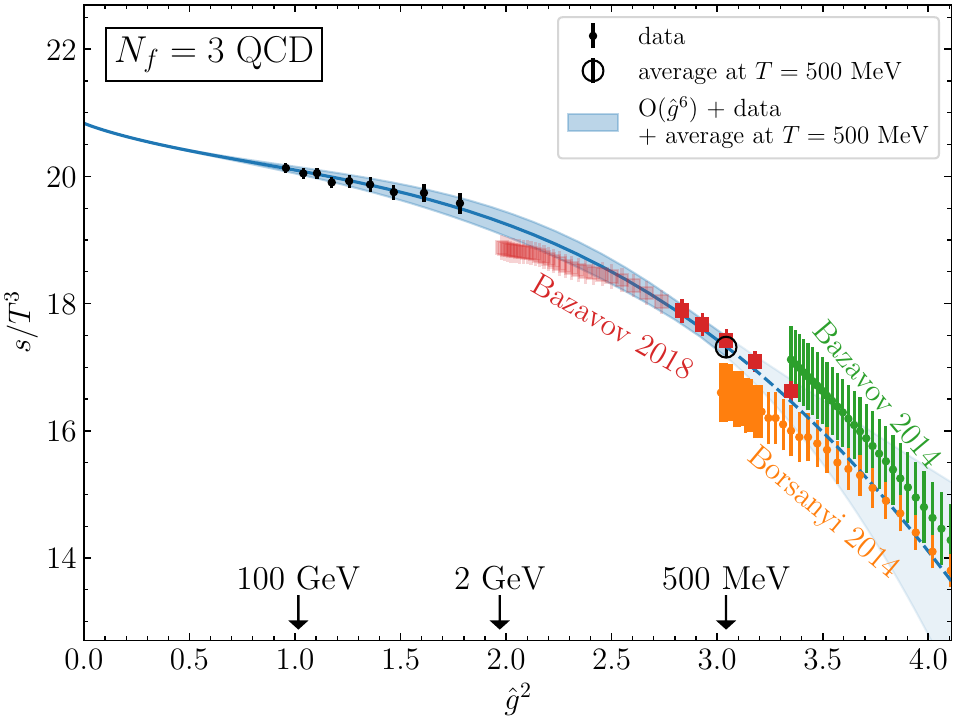}
   \end{minipage}
   \begin{minipage}{\columnwidth}
   \includegraphics[width=\textwidth]{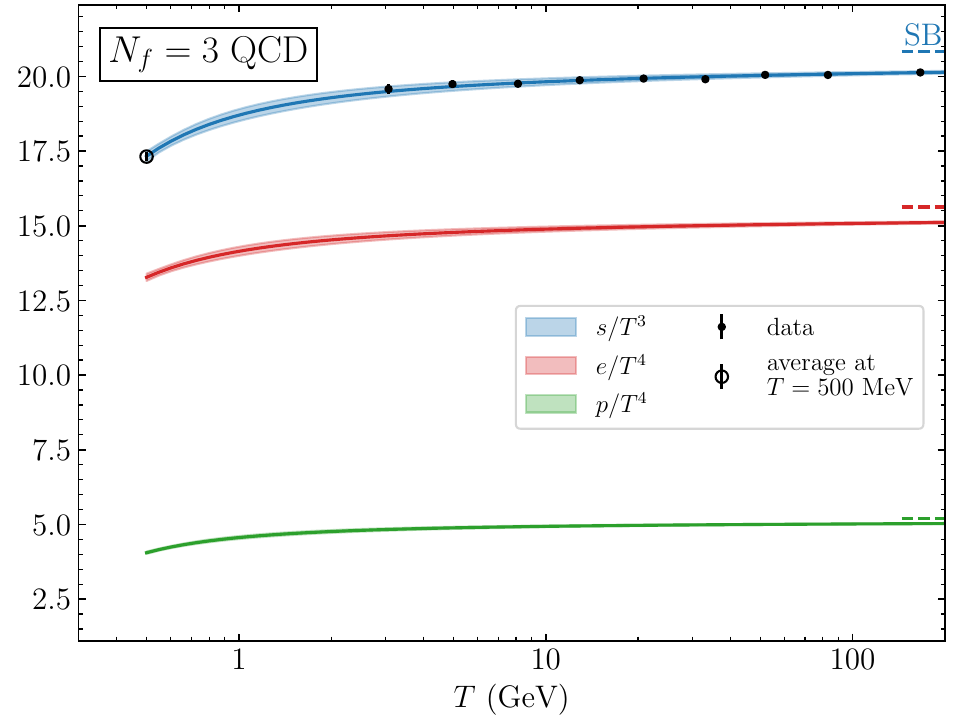}
   \end{minipage}
   \end{center}
   \caption{
   Left: the continuum results of $s/T^3$ (black points) are fitted (blue band) to the fit ansatz Eq.~\eqref{eq:fit_ansatz_2}.
   The comparison with non-perturbative results from the literature (Refs.~\cite{Borsanyi:2013bia,HotQCD:2014kol,Bazavov:2017dsy})
   is also shown.
   Right: entropy density, energy density and pressure for $T\geq 500$ MeV.
   }
   \label{fig:fitall_sep}
\end{figure*}
The results in the continuum for the entropy density are represented in the right panel of Figure~\ref{fig:clim_fitpheno} as a function of a 
renormalized coupling $\hat g(T)$, which we have set to be the five-loop $\MSbar$ coupling at the renormalization scale $\mu=2\pi T$. 
At leading order,
\begin{equation}
	\frac{1}{\hat g^2(T)} = \frac{9}{8\pi^2}\ln\left(\frac{2\pi T}{\Lambda_{\MSbar}}\right) + \cdots
\end{equation}
where $\Lambda_{\MSbar} = 341$ MeV~\cite{Bruno:2017gxd}.
For our purposes, this is only a convenient function of $\ln(T/\Lambda_\MSbar)$ which we use to parametrize the temperature dependence
of our non-perturbative data, and to facilitate the comparison with perturbation theory.
We have considered first the following phenomenological polynomial function in $\hat g$,
\begin{equation}
    \frac{s}{T^3} = \frac{32\pi^2}{45}\left[s_0 + s_2\left(\frac{\hat g}{2\pi}\right)^2 + s_3\left(\frac{\hat g}{2\pi}\right)^3\right]\,.
    \label{eq:fit_ansatz_1}
\end{equation}
A simple, linear fit in $\hat g^2$ ($s_3=0$) leads to $s_0=2.954(15)$ and $s_2=-3.6(7)$ with $\chi^2/\chi^2_{\rm exp}=0.58$. 
This fit is shown as the orange band in Figure~\ref{fig:clim_fitpheno}. The intercept is compatible within one standard
deviation with the Stefan-Boltzmann (SB) result at infinite temperature, $s_0^{\rm SB} = 2.969$.
We have then enforced $s_0=s_0^{\rm SB}$ leaving $s_2, s_3$ as fit parameters, for which we have obtained $s_2=-5.1(9)$ and $s_3=5(5)$, 
$\chi^2/\chi^2_{\rm exp}=0.56$.
This fit is represented in Figure~\ref{fig:clim_fitpheno} as the green band.
These results show that our data are well described by a simple, effective function of the strong coupling, 
pointing at the expected asymptotic limit $s_0^{\rm SB}$, in a range of temperatures that spans almost two orders of magnitude from 
3 GeV to 165 GeV. 
The resulting coefficient $s_2$ is in tension with the corresponding perturbative value $-8.438$ by a few standard deviations.

Perturbation theory is however expected to reproduce the correct thermodynamic behaviour at asymptotically high temperatures.
We have thus considered the following fit ansatz,
\begin{equation}
    \frac{s}{T^3} = \frac{32\pi^2}{45}\left[
    \sum_{k=0}^6 s_k\left(\frac{\hat g}{2\pi}\right)^k 
    + q_c\left(\frac{\hat g}{2\pi}\right)^6 + s_7\left(\frac{\hat g}{2\pi}\right)^7
    \right]\,,
    \label{eq:fit_ansatz_2}
\end{equation}
where we have enforced all the known coefficients $s_k$, $k=1,...,6$ of the perturbative expansion~\cite{Kajantie:2002wa}.
The coefficients $q_c$ and $s_7$ are fit parameters, and contain non-perturbative contributions.
In this fit we have also included the value $s/T^3 = 17.31(16)$ at $T=500$ MeV, coming from the weighted average of the non-perturbative
results of Refs.~\cite{Borsanyi:2013bia,Bazavov:2017dsy} for $N_f=2+1$ QCD.
This is legitimate because the effect of the three lightest quark masses for $T\geq 500$ MeV is
negligible with respect to the statistical accuracy of the numerical data~\cite{Laine:2006cp}.
The resulting coefficients from this fit are $q_c=-4.0(1.1)\times 10^3$ and $s_7 = 7(4)\times 10^3$, with a good fit 
quality $\chi^2/\chi^2_{\rm exp}=0.79$.
It is interesting to notice that the effect of these terms still amounts to about $40\%$ of the interactions at 
the temperature $T=165$ GeV, which is the highest in our study.
This shows that, at our level of precision, the contributions beyond the known perturbative expansion, 
which include non-perturbative effects, are relevant to properly capture the QCD thermodynamics up to the electroweak scale. 
The fit of Eq.~\eqref{eq:fit_ansatz_2} is represented as the blue band in both panels of Figure~\ref{fig:fitall_sep}. 
The associated relative error is $\lesssim 1\%$ in the entire range $T\geq 500$ MeV.

Given the temperature dependence of the entropy density, we have used it to determine the pressure $p(T)$ and the energy density $e(T)$.
Similarly to Eq.~\eqref{eq:fit_ansatz_2} for the entropy, we have parametrized the pressure as 
\begin{equation}
    \frac{p}{T^4} = \frac{8\pi^2}{45}\Bigg[
    \sum_{k=0}^6 p_k\left(\frac{\hat g}{2\pi}\right)^k
    + q_c\left(\frac{\hat g}{2\pi}\right)^6 + p_7\left(\frac{\hat g}{2\pi}\right)^7\Bigg]\,,
\end{equation}
where the coefficients $p_k$, $k=1,...,6$ have been fixed to their perturbative values~\cite{Kajantie:2002wa} while $p_7$ has been 
obtained by matching to the entropy $s = dp/dT$, which gives
\begin{equation}
	p_7 = s_7 + \frac{45}{16}\,p_5 + 3\,p_3\,.
\end{equation}
Finally, the energy density follows from $e=Ts-p$.
The complete result for the EoS of $N_f=3$ QCD for $T\geq 500$ MeV is represented in the right panel of Figure~\ref{fig:fitall_sep}, 
which shows the three thermodynamic quantities as functions of the temperature.

\section{Conclusions}
In Ref.~\cite{Bresciani:2025vxw}, and in these proceedings, we have presented the 
first non-perturbative computation of the EoS of QCD in the temperature range $3-165$ GeV.
Our primary observable is the entropy density, which we have determined with a final relative error of $0.5\%-1.0\%$.
The results show a simple and smooth effective behaviour in the strong coupling, pointing to the Stefan-Boltzmann limit
at infinite temperature.
By enforcing all the known perturbative expansion at asymptotically high temperatures, and by including 
one point at $T=500$ MeV from the literature~\cite{Borsanyi:2013bia,Bazavov:2017dsy},
we have obtained a parametrization of the entropy density in the interval $T\geq 500$ MeV.
The terms beyond the known perturbative expansion, which contain non-perturbative 
effects and which we have fitted to the numerical data, give a relevant contribution at all temperatures up to 
the electroweak scale.
This confirms that thermal QCD must be treated non-perturbatively for a reliable description of the EoS, 
a fact that has been pointed out also for other quantities like the hadronic screening masses~\cite{DallaBrida:2021ddx,Giusti:2024ohu}.

This computation is based on a new renormalization procedure for 
the lattice theory, according to which the lines of constant physics are
determined by fixing the value of the non-perturbatively defined coupling $\bar g_{\rm SF}^2(\mu)$~\cite{DallaBrida:2021ddx}.
The renormalization condition depends on the renormalization scale $\mu$, 
which can be chosen to be close to the temperature $T$ for each $T$ to be simulated.
This overcomes the technical difficulty of simulating, at the same lattice
spacing, the pion mass and the temperature, if the standard hadronic 
renormalization schemes were used.
A similar problem would arise from the zero temperature subtraction
entering in the renormalization of the thermodynamic potentials.
By considering QCD in the presence of shifted boundary conditions~\cite{Giusti:2010bb,Giusti:2011kt,Giusti:2012yj,DallaBrida:2020gux}, 
the entropy density can be directly computed without any subtraction.
All in all, thanks to these theoretical advances, 
the temperature $T$ is the relevant energy scale in our lattices
and simulations can be performed up to the electroweak scale with systematic
effects under control.

The results presented here are for QCD with $\Nf=3$ flavours.
However, the strategy is general and can be extended to 
QCD with five (massive) flavours with no further conceptual issues.

\section*{Acknowledgements}
We acknowledge PRACE for awarding us access to the HPC system MareNostrum4 
at the Barcelona Supercomputing Center (Proposals n. 2018194651 and 2021240051)
where some of the numerical results presented in this paper have been obtained. 
We also thank CINECA for providing us with a very generous access to Leonardo 
during the early phases of operations of the machine and for the computer time 
allocated via the CINECA-INFN, CINECA-Bicocca agreements. 
The R\&D has been carried out on the PC clusters Wilson and Knuth at 
Milano-Bicocca. 
We thank all these institutions for the technical support. 
This work is (partially) supported by ICSC – Centro Nazionale di Ricerca in High
Performance Computing, Big Data and Quantum Computing, funded by European 
Union – NextGenerationEU.


\begin{thebibliography}{10}
\expandafter\ifx\csname url\endcsname\relax
  \def\url#1{\texttt{#1}}\fi
\expandafter\ifx\csname urlprefix\endcsname\relax\def\urlprefix{URL }\fi
\expandafter\ifx\csname href\endcsname\relax
  \def\href#1#2{#2} \def\path#1{#1}\fi

\bibitem{Saikawa:2018rcs}
K.~Saikawa, S.~Shirai, {Primordial gravitational waves, precisely: The role of
  thermodynamics in the Standard Model}, JCAP 05 (2018) 035.
\newblock \href {http://arxiv.org/abs/1803.01038} {\path{arXiv:1803.01038}},
  \href {https://doi.org/10.1088/1475-7516/2018/05/035}
  {\path{doi:10.1088/1475-7516/2018/05/035}}.

\bibitem{Saikawa:2020swg}
K.~Saikawa, S.~Shirai, {Precise WIMP Dark Matter Abundance and Standard Model
  Thermodynamics}, JCAP 08 (2020) 011.
\newblock \href {http://arxiv.org/abs/2005.03544} {\path{arXiv:2005.03544}},
  \href {https://doi.org/10.1088/1475-7516/2020/08/011}
  {\path{doi:10.1088/1475-7516/2020/08/011}}.

\bibitem{Borsanyi:2016ksw}
S.~Borsanyi, et~al., {Calculation of the axion mass based on high-temperature
  lattice quantum chromodynamics}, Nature 539~(7627) (2016) 69--71.
\newblock \href {http://arxiv.org/abs/1606.07494} {\path{arXiv:1606.07494}},
  \href {https://doi.org/10.1038/nature20115} {\path{doi:10.1038/nature20115}}.

\bibitem{GrillidiCortona:2015jxo}
G.~Grilli~di Cortona, E.~Hardy, J.~Pardo~Vega, G.~Villadoro, {The QCD axion,
  precisely}, JHEP 01 (2016) 034.
\newblock \href {http://arxiv.org/abs/1511.02867} {\path{arXiv:1511.02867}},
  \href {https://doi.org/10.1007/JHEP01(2016)034}
  {\path{doi:10.1007/JHEP01(2016)034}}.

\bibitem{Borsanyi:2013bia}
S.~Borsanyi, Z.~Fodor, C.~Hoelbling, S.~D. Katz, S.~Krieg, K.~K. Szabo, {Full
  result for the QCD equation of state with 2+1 flavors}, Phys. Lett. B 730
  (2014) 99--104.
\newblock \href {http://arxiv.org/abs/1309.5258} {\path{arXiv:1309.5258}},
  \href {https://doi.org/10.1016/j.physletb.2014.01.007}
  {\path{doi:10.1016/j.physletb.2014.01.007}}.

\bibitem{HotQCD:2014kol}
A.~Bazavov, et~al., {Equation of state in ( 2+1 )-flavor QCD}, Phys. Rev. D 90
  (2014) 094503.
\newblock \href {http://arxiv.org/abs/1407.6387} {\path{arXiv:1407.6387}},
  \href {https://doi.org/10.1103/PhysRevD.90.094503}
  {\path{doi:10.1103/PhysRevD.90.094503}}.

\bibitem{Bazavov:2017dsy}
A.~Bazavov, P.~Petreczky, J.~H. Weber, {Equation of State in 2+1 Flavor QCD at
  High Temperatures}, Phys. Rev. D 97~(1) (2018) 014510.
\newblock \href {http://arxiv.org/abs/1710.05024} {\path{arXiv:1710.05024}},
  \href {https://doi.org/10.1103/PhysRevD.97.014510}
  {\path{doi:10.1103/PhysRevD.97.014510}}.

\bibitem{Linde:1980ts}
A.~D. Linde, {Infrared Problem in Thermodynamics of the Yang-Mills Gas}, Phys.
  Lett. B 96 (1980) 289--292.
\newblock \href {https://doi.org/10.1016/0370-2693(80)90769-8}
  {\path{doi:10.1016/0370-2693(80)90769-8}}.

\bibitem{Kajantie:2002wa}
K.~Kajantie, M.~Laine, K.~Rummukainen, Y.~Schroder, {The Pressure of hot QCD up
  to $g^6 \ln(1/g)$}, Phys. Rev. D 67 (2003) 105008.
\newblock \href {http://arxiv.org/abs/hep-ph/0211321}
  {\path{arXiv:hep-ph/0211321}}, \href
  {https://doi.org/10.1103/PhysRevD.67.105008}
  {\path{doi:10.1103/PhysRevD.67.105008}}.

\bibitem{Bresciani:2025vxw}
M.~Bresciani, M.~D. Brida, L.~Giusti, M.~Pepe, {QCD Equation of State with
  $N_f=3$ Flavors up to the Electroweak Scale}, Phys. Rev. Lett. 134~(20)
  (2025) 201904.
\newblock \href {http://arxiv.org/abs/2501.11603} {\path{arXiv:2501.11603}},
  \href {https://doi.org/10.1103/PhysRevLett.134.201904}
  {\path{doi:10.1103/PhysRevLett.134.201904}}.

\bibitem{Luscher:1991wu}
M.~L{\"u}scher, P.~Weisz, U.~Wolff, {A numerical method to compute the running
  coupling in asymptotically free theories}, Nucl. Phys. B359 (1991) 221--243.
\newblock \href {https://doi.org/10.1016/0550-3213(91)90298-C}
  {\path{doi:10.1016/0550-3213(91)90298-C}}.

\bibitem{Luscher:1993gh}
M.~L{\"u}scher, R.~Sommer, P.~Weisz, U.~Wolff, {A precise determination of the
  running coupling in the SU(3) Yang-Mills theory}, Nucl. Phys. B413 (1994)
  481--502.
\newblock \href {http://arxiv.org/abs/hep-lat/9309005}
  {\path{arXiv:hep-lat/9309005}}, \href
  {https://doi.org/10.1016/0550-3213(94)90629-7}
  {\path{doi:10.1016/0550-3213(94)90629-7}}.

\bibitem{Jansen:1995ck}
K.~Jansen, et~al., {Non-perturbative renormalization of lattice QCD at all
  scales}, Phys. Lett. B372 (1996) 275--282.
\newblock \href {http://arxiv.org/abs/hep-lat/9512009}
  {\path{arXiv:hep-lat/9512009}}, \href
  {https://doi.org/10.1016/0370-2693(96)00075-5}
  {\path{doi:10.1016/0370-2693(96)00075-5}}.

\bibitem{DallaBrida:2021ddx}
M.~Dalla~Brida, L.~Giusti, T.~Harris, D.~Laudicina, M.~Pepe, {Non-perturbative
  thermal QCD at all temperatures: the case of mesonic screening masses}, JHEP
  04 (2022) 034.
\newblock \href {http://arxiv.org/abs/2112.05427} {\path{arXiv:2112.05427}},
  \href {https://doi.org/10.1007/JHEP04(2022)034}
  {\path{doi:10.1007/JHEP04(2022)034}}.

\bibitem{Giusti:2010bb}
L.~Giusti, H.~B. Meyer, {Thermal momentum distribution from path integrals with
  shifted boundary conditions}, Phys. Rev. Lett. 106 (2011) 131601.
\newblock \href {http://arxiv.org/abs/1011.2727} {\path{arXiv:1011.2727}},
  \href {https://doi.org/10.1103/PhysRevLett.106.131601}
  {\path{doi:10.1103/PhysRevLett.106.131601}}.

\bibitem{Giusti:2011kt}
L.~Giusti, H.~B. Meyer, {Thermodynamic potentials from shifted boundary
  conditions: the scalar-field theory case}, JHEP 11 (2011) 087.
\newblock \href {http://arxiv.org/abs/1110.3136} {\path{arXiv:1110.3136}},
  \href {https://doi.org/10.1007/JHEP11(2011)087}
  {\path{doi:10.1007/JHEP11(2011)087}}.

\bibitem{Giusti:2012yj}
L.~Giusti, H.~B. Meyer, {Implications of Poincare symmetry for thermal field
  theories in finite-volume}, JHEP 01 (2013) 140.
\newblock \href {http://arxiv.org/abs/1211.6669} {\path{arXiv:1211.6669}},
  \href {https://doi.org/10.1007/JHEP01(2013)140}
  {\path{doi:10.1007/JHEP01(2013)140}}.

\bibitem{DallaBrida:2020gux}
M.~Dalla~Brida, L.~Giusti, M.~Pepe, {Non-perturbative definition of the QCD
  energy-momentum tensor on the lattice}, JHEP 04 (2020) 043.
\newblock \href {http://arxiv.org/abs/2002.06897} {\path{arXiv:2002.06897}},
  \href {https://doi.org/10.1007/JHEP04(2020)043}
  {\path{doi:10.1007/JHEP04(2020)043}}.

\bibitem{Laine:2006cp}
M.~Laine, Y.~Schroder, {Quark mass thresholds in QCD thermodynamics}, Phys.
  Rev. D 73 (2006) 085009.
\newblock \href {http://arxiv.org/abs/hep-ph/0603048}
  {\path{arXiv:hep-ph/0603048}}, \href
  {https://doi.org/10.1103/PhysRevD.73.085009}
  {\path{doi:10.1103/PhysRevD.73.085009}}.

\bibitem{Wilson:1974sk}
K.~G. Wilson, {Confinement of Quarks}, Phys. Rev. D 10 (1974) 2445--2459.
\newblock \href {https://doi.org/10.1103/PhysRevD.10.2445}
  {\path{doi:10.1103/PhysRevD.10.2445}}.

\bibitem{Wilson:1975hf}
K.~G. Wilson, {Quarks: From Paradox to Myth}, Subnucl. Ser. 13 (1977) 13--32.

\bibitem{Sheikholeslami:1985ij}
B.~Sheikholeslami, R.~Wohlert, {Improved Continuum Limit Lattice Action for QCD
  with Wilson Fermions}, Nucl. Phys. B259 (1985) 572.
\newblock \href {https://doi.org/10.1016/0550-3213(85)90002-1}
  {\path{doi:10.1016/0550-3213(85)90002-1}}.

\bibitem{Yamada:2004ja}
N.~Yamada, et~al., {Non-perturbative O(a)-improvement of Wilson quark action in
  three-flavor QCD with plaquette gauge action}, Phys. Rev. D71 (2005) 054505.
\newblock \href {http://arxiv.org/abs/hep-lat/0406028}
  {\path{arXiv:hep-lat/0406028}}, \href
  {https://doi.org/10.1103/PhysRevD.71.054505}
  {\path{doi:10.1103/PhysRevD.71.054505}}.

\bibitem{Luscher:1992an}
M.~L{\"u}scher, R.~Narayanan, P.~Weisz, U.~Wolff, {The Schr{\"o}dinger
  functional: a renormalizable probe for non-Abelian gauge theories}, Nucl.
  Phys. B384 (1992) 168--228.
\newblock \href {http://arxiv.org/abs/hep-lat/9207009}
  {\path{arXiv:hep-lat/9207009}}, \href
  {https://doi.org/10.1016/0550-3213(92)90466-O}
  {\path{doi:10.1016/0550-3213(92)90466-O}}.

\bibitem{Bruno:2017gxd}
M.~Bruno, M.~Dalla~Brida, P.~Fritzsch, T.~Korzec, A.~Ramos, S.~Schaefer,
  H.~Simma, S.~Sint, R.~Sommer, {QCD Coupling from a Nonperturbative
  Determination of the Three-Flavor $\Lambda$ Parameter}, Phys. Rev. Lett.
  119~(10) (2017) 102001.
\newblock \href {http://arxiv.org/abs/1706.03821} {\path{arXiv:1706.03821}},
  \href {https://doi.org/10.1103/PhysRevLett.119.102001}
  {\path{doi:10.1103/PhysRevLett.119.102001}}.

\bibitem{DallaBrida:2018rfy}
M.~Dalla~Brida, P.~Fritzsch, T.~Korzec, A.~Ramos, S.~Sint, R.~Sommer, {A
  non-perturbative exploration of the high energy regime in $N_{\mathrm{f}}=3$
  QCD}, Eur. Phys. J. C 78~(5) (2018) 372.
\newblock \href {http://arxiv.org/abs/1803.10230} {\path{arXiv:1803.10230}},
  \href {https://doi.org/10.1140/epjc/s10052-018-5838-5}
  {\path{doi:10.1140/epjc/s10052-018-5838-5}}.

\bibitem{inprep}
M.~Bresciani, M.~Dalla~Brida, L.~Giusti, M.~Pepe, {Publication in preparation}.

\bibitem{Bruno:2022mfy}
M.~Bruno, R.~Sommer, {On fits to correlated and auto-correlated data}, Comput.
  Phys. Commun. 285 (2023) 108643.
\newblock \href {http://arxiv.org/abs/2209.14188} {\path{arXiv:2209.14188}},
  \href {https://doi.org/10.1016/j.cpc.2022.108643}
  {\path{doi:10.1016/j.cpc.2022.108643}}.

\bibitem{Giusti:2024ohu}
L.~Giusti, T.~Harris, D.~Laudicina, M.~Pepe, P.~Rescigno, {Baryonic screening
  masses in QCD at high temperature}, Phys. Lett. B 855 (2024) 138799.
\newblock \href {http://arxiv.org/abs/2405.04182} {\path{arXiv:2405.04182}},
  \href {https://doi.org/10.1016/j.physletb.2024.138799}
  {\path{doi:10.1016/j.physletb.2024.138799}}.

\end{thebibliography}
\end{document}